\documentclass[compsoc]{IEEEtran}
\ifCLASSOPTIONcompsoc
  \usepackage[nocompress]{cite}
\else
  \usepackage{cite}
\fi
\ifCLASSINFOpdf
  \usepackage[pdftex]{graphicx}
  \DeclareGraphicsExtensions{.pdf,.jpeg,.png}
\else
  \usepackage[dvips]{graphicx}
  \DeclareGraphicsExtensions{.eps}
\fi
\usepackage{amsmath}

\usepackage{latexsym,bm}
\usepackage{color}

\begin{document}
\title{PEPS++: Towards Extreme-scale Simulations of Strongly Correlated Quantum Many-particle Models on Sunway TainhuLight
}


% compsoc conferences compared to traditional conferences), use this
\author{\IEEEauthorblockN{
Lixin He\IEEEauthorrefmark{1},
Hong An\IEEEauthorrefmark{2},
Chao Yang\IEEEauthorrefmark{3}\IEEEauthorrefmark{4},
Fei Wang\IEEEauthorrefmark{5},
Junshi Chen\IEEEauthorrefmark{2},
Chao Wang\IEEEauthorrefmark{1},
Weihao Liang\IEEEauthorrefmark{2},\\
Shaojun Dong\IEEEauthorrefmark{1},
Qiao Sun\IEEEauthorrefmark{3},
Wenting Han\IEEEauthorrefmark{2},
Wenyuan Liu\IEEEauthorrefmark{1},
Yongjian Han\IEEEauthorrefmark{1} and
Wenjun Yao\IEEEauthorrefmark{2}
}\\
\IEEEauthorblockA{\IEEEauthorrefmark{1}
CAS Key Lab of Quantum Information,
University of Science and Technology of China,
Hefei, China
}\\
\IEEEauthorblockA{\IEEEauthorrefmark{2}
School of Computer Science \& Technology,
University of Science and Technology of China,
Hefei, China
}\\
\IEEEauthorblockA{\IEEEauthorrefmark{3}
CCSE \& CAPT, School of Mathematical Sciences \& NELVT, Peking University, Beijing, China
}\\
\IEEEauthorblockA{\IEEEauthorrefmark{4}
Institute of Software \& State Key Laboratory of Computer Science, CAS, Beijing, China
}\\
\IEEEauthorblockA{\IEEEauthorrefmark{5}
National Research Center of Parallel Computer Engineering and Technology, Beijing, China
}
}


\maketitle

\begin{abstract}
The study of strongly frustrated magnetic systems has drawn great attentions from both theoretical and experimental physics. Efficient simulations of these models are essential for understanding their exotic properties. Here we present PEPS++, a novel computational paradigm for simulating  frustrated magnetic systems and other strongly correlated quantum many-body systems. PEPS++ can accurately solve these models at the extreme scale with low cost and high scalability on modern heterogeneous supercomputers. We implement PEPS++ on Sunway TaihuLight based on a carefully designed tensor computation library for manipulating high-rank tensors and optimize it by invoking various high-performance matrix and tensor operations. By solving a 2D strongly frustrated $J_1$-$J_2$ model with over ten million cores, PEPS++ demonstrates the capability of simulating strongly correlated quantum many-body problems at unprecedented scales with accuracy and time-to-solution far beyond the previous state of the art.
%, thus bringing the faithful simulation of the fascinating QSL states into the reality.
\end{abstract}


% \fi
\IEEEpeerreviewmaketitle

\section{Introduction}
%Ever since Nobel Prize laureate P. W. Anderson proposed that the resonating valance-bond (RVB) state,
%a special QSL,
%might underlie the physics of high-temperature superconductivity~\cite{Anderson1987},
%there have been extensive studies on the QSL.
%% both theoretically and experimentally.
%Experimentally, QSL has been searched mostly in
%layered frustrated magnetic materials, e.g., Herbertsmithite \cite{Norman2016}, whereas
%the theoretical studies on QSL are focused on various two-dimensional (2D) models, including
%$J_1$-$J_2$ model (see Fig.\ref{fig:QSL}), kagome model and Kitaev model etc. \cite{Savary2017}.

The numerical methods play the key role for us to understand the physics from the cosmology to many-body physics. Even though, it have obtained great achievements, to find an effective numerical method to attack the 2D strongly correlated and frustrated models is still a big challenge. The popular density functional theory (DFT) \cite{Kohn1999,Gygi2006,Hasegawa2011}, and other
perturbative methods, although have made great successes for
simulating weakly correlated material systems,
will completely fail for the strongly correlated systems.
The dynamical mean field theory \cite{Georges1996} that is effective for 3D systems
is also unfit for 2D systems which have stronger quantum fluctuations.
Revealing the fascinating physical natures in these systems
mainly relays on the non-perturbative numerical methods,
such as exact diagonalization (ED),
quantum Monte Carlo (QMC) method \cite{Anderson2007}  and density matrix renormalization group (DMRG)~\cite{white92}.
%developing new efficient algorithms is still urgent, because of there are serious limitations of the previous methods:
However, these methods all have serious limitations: e.g., the computational cost of ED grows exponentially with the system size,
and therefore the system size of the ED method is limited to less than 50 sites\cite{Savary2017}.
QMC suffers from the notorious sign problem for fermionic and frustrated systems\cite{Troyer05};
and DMRG is limited to 1D or quasi-1D systems and does not work well for higher dimension systems\cite{Schollw11}. Recently, inspired by the insight of quantum entanglement in the perspective of quantum information theory,
a class of wave functions, namely tensor network states (TNS) are proposed to describe the  strongly correlated systems.
Especially in 2D, the projected entangled pair states (PEPS)\cite{verstraete04,verstraete06}
have been proved to be powerful simulation methods to explore the strongly correlated systems.
The tensor network states methods open the possible way to understand the 2D strongly correlated physics.

Considering a 2D square lattice with $N$=$L_x\times L_y$ sites, and
$d$-dimensional local Hilbert space (which we call ``spin'' in this work),
denoted as $|s_m\rangle$ on the site $m$=($i$,$j$),
the PEPS wave function of this system can be written as\cite{verstraete04},
\begin{equation}
  |\Psi_{\rm PEPS}\rangle  = \sum_{s_m=1}^d { \rm Tr}  (A_{1}^{s_{1}} A_{2}^{s_{2}}
  \cdots A_{N}^{s_{N}})  |s_{1} \cdots s_{N} \rangle,
   \label{Eq:PEPS}
\end{equation}
where $A^{s_{m}}_{m}$=$A_{m}(l,r,u,d,s_{m})$ is a
rank-five tensor located on site $m$.
The physical index $s_{m}$ takes value from $1$ to $d$
and four virtual indices $l,r,u,d$, which correspond to four nearest neighbors.
The dimension of each virtual bond is $D$. and
the ``${\rm Tr}$'' denotes the contraction over all the virtual indices of the tensor network.
PEPS can be viewed as a generation of DMRG, which is based on the matrix product states
(MPS), in higher dimensions\cite{verstraete04}.

In PEPS, the many-particle wave functions are represented by parameters which number grows up
polynomially with the size of the system instead of exponential one. The ground state can be obtained
variationally. In this sense, the PEPS (more generally the TNS) method is very similar to
the variational quantum Monte Carlo (VQMC) method, which is well established and
has been applied successfully
to many strongly-correlated systems\cite{Martin2016,Gubernatis2016}.
Unlike the traditional VQMC method, which rely strongly on the trial wavefunctions,
PEPS is in-principle an exact method, provided the virtual bond dimension $D$
is large enough and can systematically approach the exactly desired state.

However, there are two major difficulties that impede the efficiency and the applications of the PEPS.
First, the computational costs have extremely high scaling to $D$.
For example, to contract a PEPS, the computation cost is at least
O$(D^{10})$ for the open boundary conditions (OBC) with nearest-neighbor interactions,
and O$(D^{18})$ for the periodic boundary conditions (PBC).
Furthermore, because of the strongly coupled nature of PEPS, extremely high data throughput and global communications
are required, which makes the parallelization of PEPS very difficult,
especially on today's heterogeneous supercomputers.
No massively parallelization of the PEPS
has been reported so far, and, without significant innovations, is unlikely to be achieved in the future.
As a result, the bond dimension $D$ in most existing simulations were restricted to small values, which in turn limits
the application of this powerful method in real applications.

%
%\blue{
%To reach a real two (not quasi-one) dimension physics,
%PEPS was proposed to represent the ground state wave function of the 2D quantum systems  in Ref.~\cite{verstraete04}, which shows many advantages
%over other methods.
%Originally, the ground state was obtained by optimizing
%the energy of the system for square lattice with open boundary condition (OBC)~\cite{verstraete04,verstraete08};
%latter, an imaginary time evolution method was implemented to
%optimize the PEPS wave functions in a translational invariant system\cite{jordan08}.
%These methods obtained beautiful results for the Ising model. However, the computational cost scales as high as O$(D^{10})$.
%Such high computational cost limits the bond dimension $D$ to quite small values, e.g., $D$=4 in Ref.\cite{verstraete04},
%and as a result, the application of the method is rather limited.}

To reduce the calculation cost, an approximate imaginary time evolution with simple update (SU) algorithm was proposed~\cite{simpleupdate}.
In this scheme, the environment of a tensor is approximated by products of some diagonal matrices, and therefore it
substantially reduces the calculation cost in the update process to O$(D^5)$ for the nearest neighbor (NN) interactions~\cite{wang11}.
%Wang et al.~\cite{wang11}
%%proposed simulating periodic spin systems using PEPS combing tensor renormalization group method.
%A PEPS calculation up to $D$=9
%%and system size $24 \times 24$
%has been demonstrated~\cite{wang11}.
Unfortunately, this approximation brings some uncontrolled errors into the calculations, and therefore
the accuracy of these calculations is quite low~\cite{wang16,lubasch14}.
% Even for very small systems,
%e.g., a $6\times 6$ lattice, the absolute error is as high as 0.001~\cite{wang16}.
%A even larger $D$ has been reached ($D$=25) in the kagome lattice with SU(2) symmetry (arxiv:1606.09639)
%using the SU scheme.
%Because the effects of environment are over simplified in the SU methods,
%the PEPS may not converge to the ground state of the system with desired precision~\cite{lubasch14}
%and the corresponding calculated physical properties may not be reliable.
%Furthermore, after the optimization of PEPS, the calculations of physical quantities still scales
%as high as O($D^{10}$), unless further approximations are made.

Great effort has been made to improve the accuracy of SU
~\cite{orus15,singlelayer}, e.g. the cluster update method\cite{lubasch14} and the full update (FU) method \cite{lubasch14,fullupdate14}, etc. which taken the enviroment into better account.
%Recently a cluster update method\cite{lubasch14}, allowing a tradeoff between computation cost and precision, is proposed to improve the accuracy systematically by approximating the environment of local tensors with different clusters of sizes. Based on the cluster update method, a full update (FU) method \cite{lubasch14,fullupdate14}, meaning taking the whole lattice into account, can significantly improve the accuracy from that of the SU.
Unfortunately, the computational scaling of the FU is still O$(D^{10})$ for the nearest neighbor interactions,
which prevents one using larger $D$ in PEPS further.
For many strongly correlated systems which are highly entangled with rather long correlation length,
large bond dimension $D$ is critical to get reliable results.

%However, because of the expensive cost, the largest bond dimension has been used by the full update method so far
%is $D$=7 by taking the advantages of SU(2) symmetry. It seems
%unlikely to go to a larger $D$ with current method.
%\blue{To summarize, so far the application of PEPS is still at its early stage.
%Accurate calculations based on PEPS are limited to very small $D$ (generally $D\leq$7,
%even by taking the advantages of symmetries, which are not always applicable),
%and small system size (less than 10$\times$10 non-equivalent sites).
%With uncontrolled SU approximations, one can go to much larger $D$ ($<$25)\cite{PhysRevLett.118.137202}
%and system size ($\sim$ 24$\times$24 or
%infinite system with translation symmetry),
%but the accuracy of the methods may not be enough for the most interesting highly entangled many-particle states.
%%There are two major difficulties of the method that prevent the use of large $D$.
%%Besides the the extremely high cost scaling to the virtual
%%bond dimension $D$, the PEPS method is intrinsically very difficult to parallelize,
%%even more difficult than the closely related DMRG method,
%%because of their smaller tensor dimensions, and complex and irregular data structures.
%}

In this work, we present a novel low scaling and highly scalable method, which takes the advantages
of both PEPS and Monte Carlo (MC) techniques, to simulate the strongly correlated
and frustrated 2D quantum systems on many-core based massively parallel clusters at the extreme scale.
We note that by taking the advantages of symmetries, e.g., SU(2)
\cite{PhysRevX.2.041013,PhysRevLett.113.046402}, one can significantly reduce the computational costs. More specifically, for a translation invariant system (where only small unite cell is considered), one can
treat infinite systems directly via so called infinite PEPS (iPEPS)\cite{PhysRevLett.118.137202,Corboz2016,PhysRevX.2.041013,PhysRevLett.113.046402}. However, these symmetries are not always applicable.
Therefore, here we focus on the most general case on finite size systems, where no symmetries are enforced.
In this algorithm, the energy together with energy gradients of a given PEPS
are calculated through the MC sampling technique which
dramatically reduces the scaling of the calculation to O$(MD^6)$, where $M$ is the sampling sweeps,
significantly less than the original method.
The physical quantities can also be calculated using the the MC sampling technique, with the same computational scaling.
This new method, which is called PEPS++, is intrinsically parallel,
and therefore paves a new path for us to investigate the strongly correlated quantum systems
e.g., quantum spin liquids at unprecedented scale with controlled accuracy.

We demonstrate our methods on the frustrated $J_1$-$J_2$ model, which
is considered as one of the most challenging ones. The frustrated interactions ($J_2$) result in a
large degeneracy of the ground state, and the quantum fluctuation may lead to massive coherent superposition of the
degenerated states, implying a novel highly entangled (correlated)
quantum state, known as quantum spin liquid (QSL)~\cite{Anderson1973,Anderson1987,Balents2010},
which lacks any long range magnetic order even down to zero temperature. Because of the anomalously high degree of entanglement, QSLs have nontrivial topological properties which may host exotic excitations with fractional statistics, such as spinons, and visions, etc.,
which have important applications in quantum computing\cite{Kitaev2003,Kitaev2006}.
It is widely believed that this peculiar ground state will appear in the region near $J_2 \sim$0.5$J_1$.
However, the exact nature of the model near the $J_2 \simeq$0.5$J_1$ region
is still controversially unclear.
%Consequently, this problem could only be solved by approximate methods in the past.
Because of the frustrated character, the ground state will possess long correlation length and high entanglement, large bond dimension and system size
are both eagerly required to bring reliable simulations into the reality.
By demonstrating our method, named as PEPS++, on the frustrated $J_1$-$J_2$ model, especially for
the fascinating $J_2 \sim$0.5$J_1$ region, we successfully obtain
accurate simulation results at an unprecedented scale of entangled 24$\times$24 sites
by using over 10 million heterogeneous cores from the world-leading Sunway TaihuLight supercomputer,
and thus enable a promising paradigm for solving this long standing model in quantum physics.

%\section{Related Work}

\section{Innovations Realized}

To deal with the severe challenges arising from strongly correlated quantum many-body systems,
we need an ingenious strategy that is highly distinct from all above mentioned methods.
To that end, we propose PEPS++, which is a novel computational paradigm with three major advantages listed as follows.
\begin{itemize}
\item High scalability: The method combines PEPS with MC sampling, therefore can
provide two levels of parallelism and is hardware friendly on modern heterogeneous supercomputers.
\item Low cost: The computational complexity of the proposed method is O$(M D^6)$, therefore is
able to handle systems with very large bond dimension $D$ far beyond the current state of the art.
\item Controlled accuracy: Unlike many approximate algorithms, the proposed method can deliver
faithful simulation results on challenging strongly correlated systems with the help of
a gradient optimization technique.
\end{itemize}
We summarize below the main aspects of PEPS++, and show how it is implemented and optimized on the Sunway TaihuLight
supercomputer to bring unprecedented computing capability into full play.

\begin{figure}
 \centering
 \includegraphics[width=3.0in]{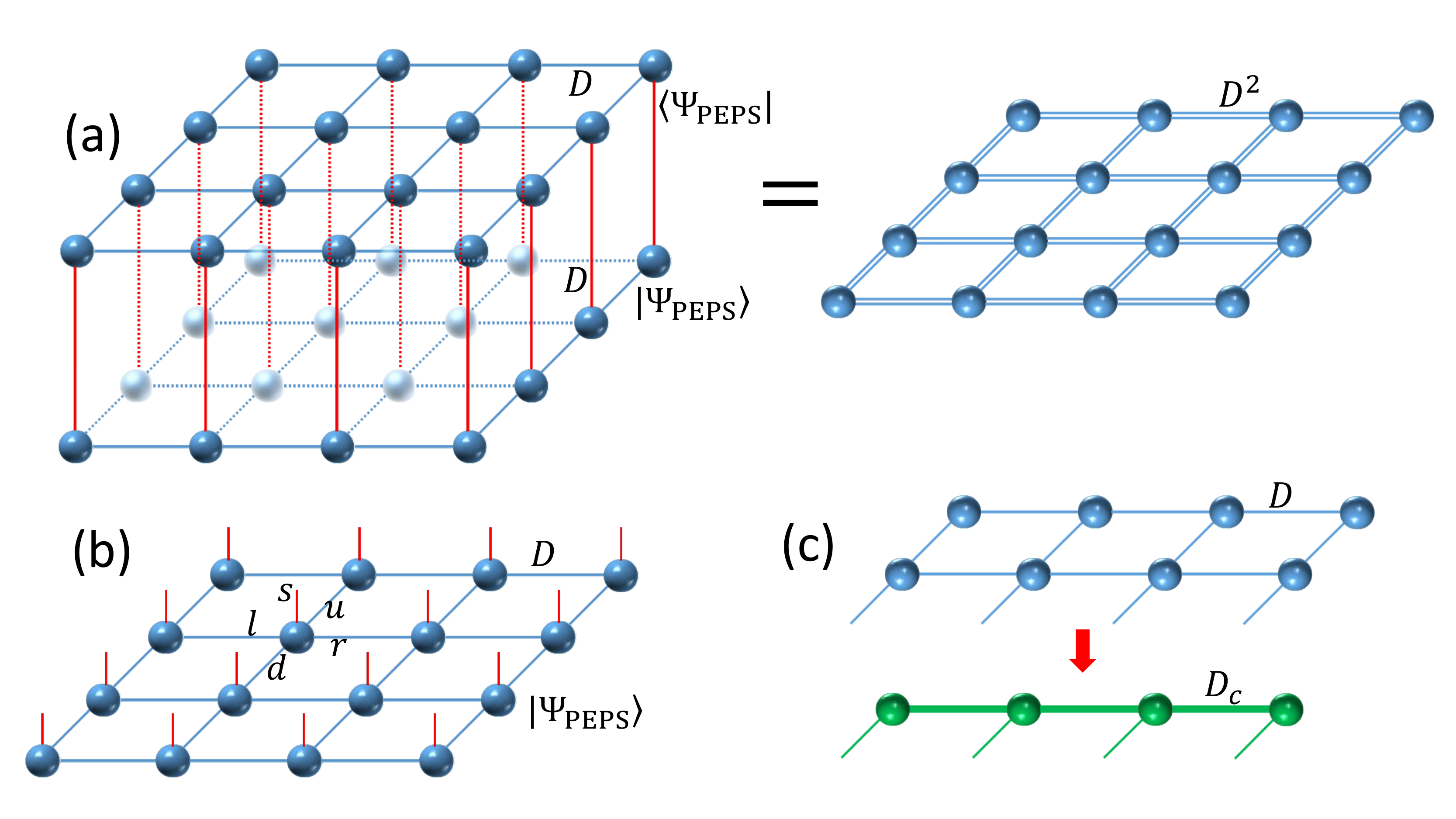}
 \caption{(a) A double-layer tensor network composed of bra $\langle \Psi_{\rm PEPS}|$ and ket $|\Psi_{\rm PEPS}\rangle$ with bond dimension $D$. After contraction
 of spin configuration $|S\rangle$, it reduces to a network with bond dimension $D^2$.
 (b) The single-layer tensor network for a given spin configuration.
 (c) When two lines of tensors are contracted, a bond dimension truncation $D_c$ is introduced.}
 \label{fig:PEPS}
\end{figure}

%\subsection{Overview of PEPS method}

\subsection{PEPS++ Algorithm}

%To extend the ability to investigate TNS of larger $D$,

The PEPS provides a systematically improvable variational wave functions for
strongly correlated quantum many-body problems.
For a given PEPS wave function $|\Psi_{\rm PEPS}\rangle$,
the total energy can be calculated as
\begin{equation}
E={\langle \Psi_{\rm PEPS}|\hat{H}|\Psi_{\rm PEPS}\rangle
\over \langle \Psi_{\rm PEPS} | \Psi_{\rm PEPS}\rangle},
\label{eq:energy}
\end{equation}
which needs to be minimized to obtain the ground state.
The PEPS variational wave functions are given by Eq.~\ref{Eq:PEPS}, where $l$, $r$, $u$, $d$=1, $\cdots$, $D$.
Here $D$ is known as the bond dimension.
To calculate the energy, we need to contract the {\it double-layer} PEPS  ($\langle \Psi_{\rm PEPS}|$, and $|\Psi_{\rm PEPS}\rangle$) i.e., to sum over all dummy variables, as schematically shown in Fig. \ref{fig:PEPS}(a).
It is easy to see that exactly contracting PEPS will lead to the exponentially growth
of the bond dimensions with the number of the lines contracted
during the process\cite{verstraete06}.
Therefore a truncation of the bond dimension
is necessary during the contraction~\cite{verstraete08}.
In the standard contraction methods to calculate Eq.~\ref{eq:energy},
the computational complexity is at least O$(D^4D_{c2}^3+dD^6D_{c2}^2)$ for OBC.
The errors induced by the bond dimension truncation can be well controlled by systematically
increasing $D_{c2}$, and usually the cut-off bond dimension $D_{c2}\propto D^2$ is enough
to bring high accuracy.
Therefore, in order to contract the whole PEPS,
the computational scaling with respect to $D$ is at least
O$(D^{10})$ for OBC with nearest-neighbor interactions only.
Because the contraction process is highly coupled, it is difficult to
parallelize.

Note that the calculation of energy in Eq.~\ref{eq:energy} can be re-expressed as follows,
\begin{equation}
  E= \frac{\langle \Psi_{\rm PEPS} |H|\Psi_{\rm PEPS}\rangle}{\langle \Psi_{\rm PEPS} | \Psi_{\rm PEPS}\rangle}=\frac{1}{Z}\sum_{S}{W^{2}(S)E(S)} ~~,
  \label{eq:energy-MC}
 \end{equation}
with
\begin{equation}
E(S)=\sum_{S^{\prime}} \frac{W(S^{\prime})}{W(S)} \langle S^{\prime}|H|S\rangle \,.
\label{eq:ess}
\end{equation}
Here $|S\rangle=|s_{1}s_{2}\cdots s_{N} \rangle$ is the
short notation for spin configuration and
\begin{equation}
W(S)= {\rm Tr}  (A_{1}^{s_{1}} A_{2}^{s_{2}}
  \cdots A_{N}^{s_{N}})  \nonumber\, .
\end{equation}
%is the weight of $|S\rangle$.
$Z=\sum_{S}W^2(S)$ is the normalization factor.
The total number of spin configurations is $d^N$, but
the energy can be evaluated through MC importance sampling according to the configuration weight $W^{2}(S)$\cite{sandvik07,schuch08}.
In the MC scheme, the most time consuming part is to calculate $W(S)$,
which is obtained by contracting a single-layer network,
rather than the double-layer one in the original method.
%PEPS with bond dimension $D$ with fixed spin configurations, shown in Fig.~\ref{fig:contract}(c).
Similarly, we need to introduce truncation $D_{c1}$ to contract PEPS, and the
cost of this process is O$(D^4D_{c1}^2)$. In high contrast
to $D_{c2}$$\sim$$D^2$ in the double-layer PEPS, $D_{c1} \propto D$
is sufficient for the single-layer PEPS,
and therefore the total computational complexity to calculate the energy is O$(D^6)$.

The energy derivation with respect to the tensor element $A_{lrud}^{s_m}$ can also be evaluated by MC sampling as:
\begin{equation}
\frac{\partial E}{\partial A_{lrud}^{s_m}}=2\langle\Delta_{lrud}^{s_m}(S)E(S)\rangle-2\langle\Delta_{lrud}^{s_m}(S)\rangle \langle E(S)\rangle \, ,
\label{eq:gradients}
\end{equation}
where $s_m$ is the physical index of tensor $A$ located on site $m$,
and $\langle \cdots \rangle$ denotes the MC average. $\Delta_{lrud}^{s_m}$ is defined as
\begin{equation}
\Delta_{lrud}^{s_m}(S)=\frac{1}{W(S)}\frac{\partial W(S)}{\partial A_{lrud}^{s_m}}=\frac{1}{W(S)} B_{lrud}^{s_m}(S) ~~ .
\label{eq:delta}
\end{equation}
In the above equation, $B_{lrud}^{s_m}(S)$ is the element of tensor
\begin{equation}
 B^{s_m}(S)={\rm Tr} (A_1^{s_1} A_2^{s_2}\cdots A_{m-1}^{s_{m-1}}A_{m+1}^{s_{m+1}}\cdots A_N^{s_N} ) ~~,
\end{equation}
which is nothing but a tensor summing over all the indices except those linked with site $m$ on the fixed configuration $|S \rangle$.
The cost of calculating the energy gradient is also O$(D^6)$.

Once we have the energy gradients, the total energy can be optimized via a gradient optimization
(GO) algorithm \cite{liu2017go}, with tensor elements updating as follows:
\begin{equation}
  A_{lrud}^{s_m}(n+1)=  A_{lrud}^{s_m}(n)-p \cdot\delta t(n)\cdot {\rm{sign}}(\frac{\partial E}{\partial A_{lrud}^{s_m}}) ~~,
\end{equation}
where $p \in$(0,1) is a random number for each $A_{lrud}^{s_m}$ and $\delta t(n)$ is the step length.
Only  the correct signs of energy derivations ($\frac{\partial E}{\partial A_{lrud}^{s_m}}$)
are used instead of the exact values in this method~\cite{sandvik07}.
The random step lengthes may help avoid the local minima.
However, our tests show that a direct GO optimization starting from a random PEPS is very expensive
and often trapped at local minima. To save the computational time,
we first perform the imaginary time evolution method with SU, a fast local optimization
method, to get a very good approximate state to the exact ground state.
The PEPS wave function is
then further optimized via the GO method.
Our results show that the combined SU method and GO method has been demonstrated to be very robust
and effective to optimize the PEPS.
%\blue{In PEPS++, all approximations are well controlled, e.g, the introduction of cut-off bond dimension $D_{c1}$ in the contraction.}

\subsection{Parallelization Strategies}

The overall computational complexity of the PEPS++ method is O($MD^6$) as compared to
O($D^{12}$) in direct contraction method for the $J_1$-$J_2$ model,
where $M$ is the MC sampling numbers.
One of the most important advantages of PEPS is that
it is  particularly suitable for
massive parallelization on modern supercomputers
with both process- and thread-level parallelisms.
The parallelization strategies are described below.

\subsubsection{Process-level parallelism}

\begin{figure}[hbt]
  \centering
  \includegraphics[width=0.45\textwidth]{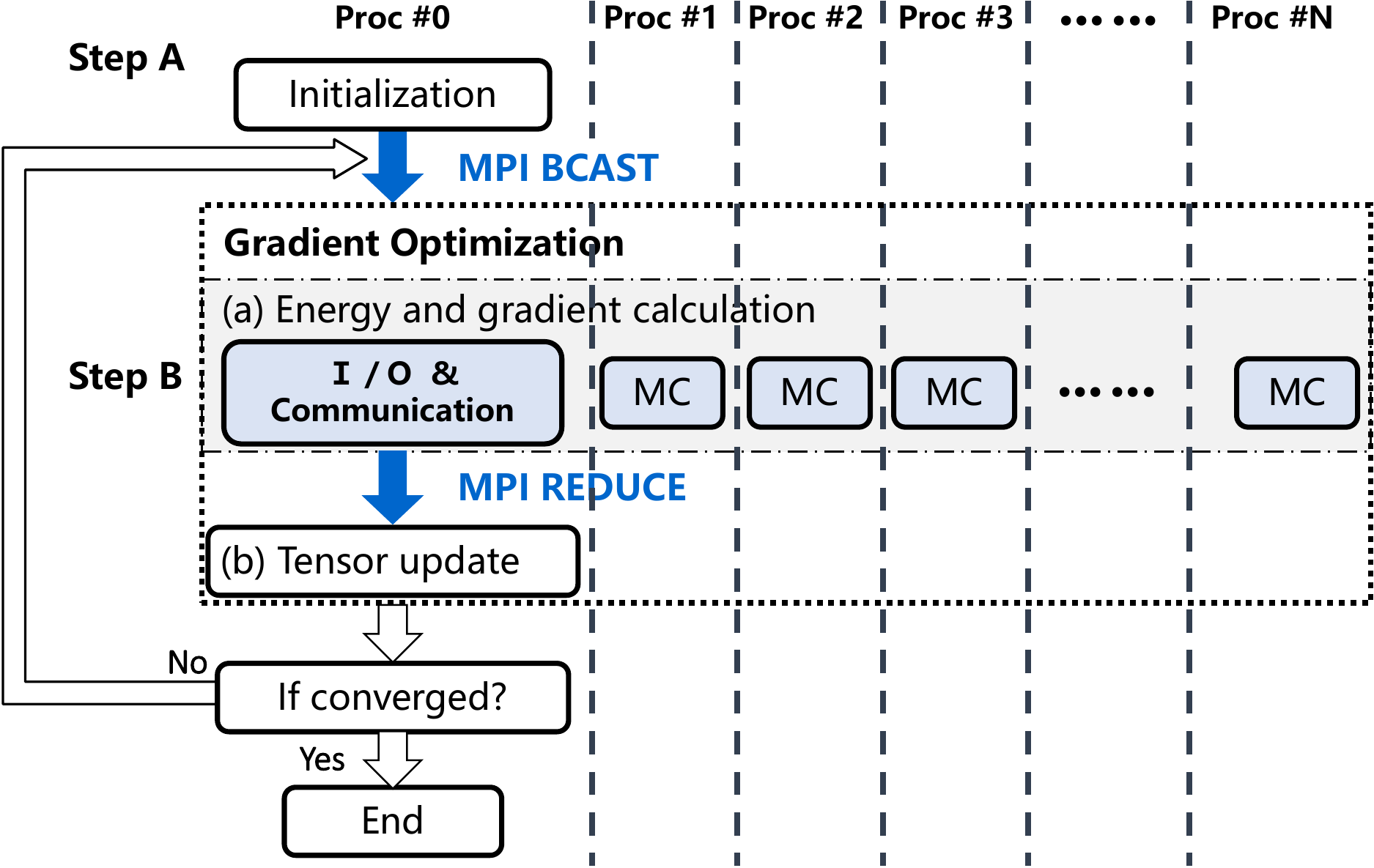}
  \caption{The process-level parallelization of PEPS++.}
  \label{fig:algorithm}
\end{figure}

The main purpose of PEPS++ is to optimize tensor elements of the tensor network to get the ground state with
the lowest energy possible.
To accomplish this goal, the program executes in steps as shown in Figure~\ref{fig:algorithm}.
The computation starts from a precomputed initial approximate state obtained from the imaginary time evolution method with SU (Step A),
which is read from a file. Then in the main loop of the gradient optimization (Step B),
the total energy and the energy gradient are computed
through Eq.~\ref{eq:energy} and Eq.~\ref{eq:gradients} respectively
in the MC sampling process (Step B-a), which takes most of the execution time,
before calculating the tensor corrections (Step B-b). The iteration of the main loop ends when the energy approaches
to a given criteria, which is chosen to be $10^{-5}$ here.
Considering the fact that the sum over spin configurations $|S\rangle$ is done
by importance sampling with independent Markov chains, which are intrinsically parallel,
we distribute the MC sampling tasks across all MPI processes to achieve good parallel scalability
except process \#0, which is solely used for data distribution and collection.

\subsubsection{Thread-level parallelism}

\begin{figure*}[htb]
  \centering
  \includegraphics[width=0.95\textwidth]{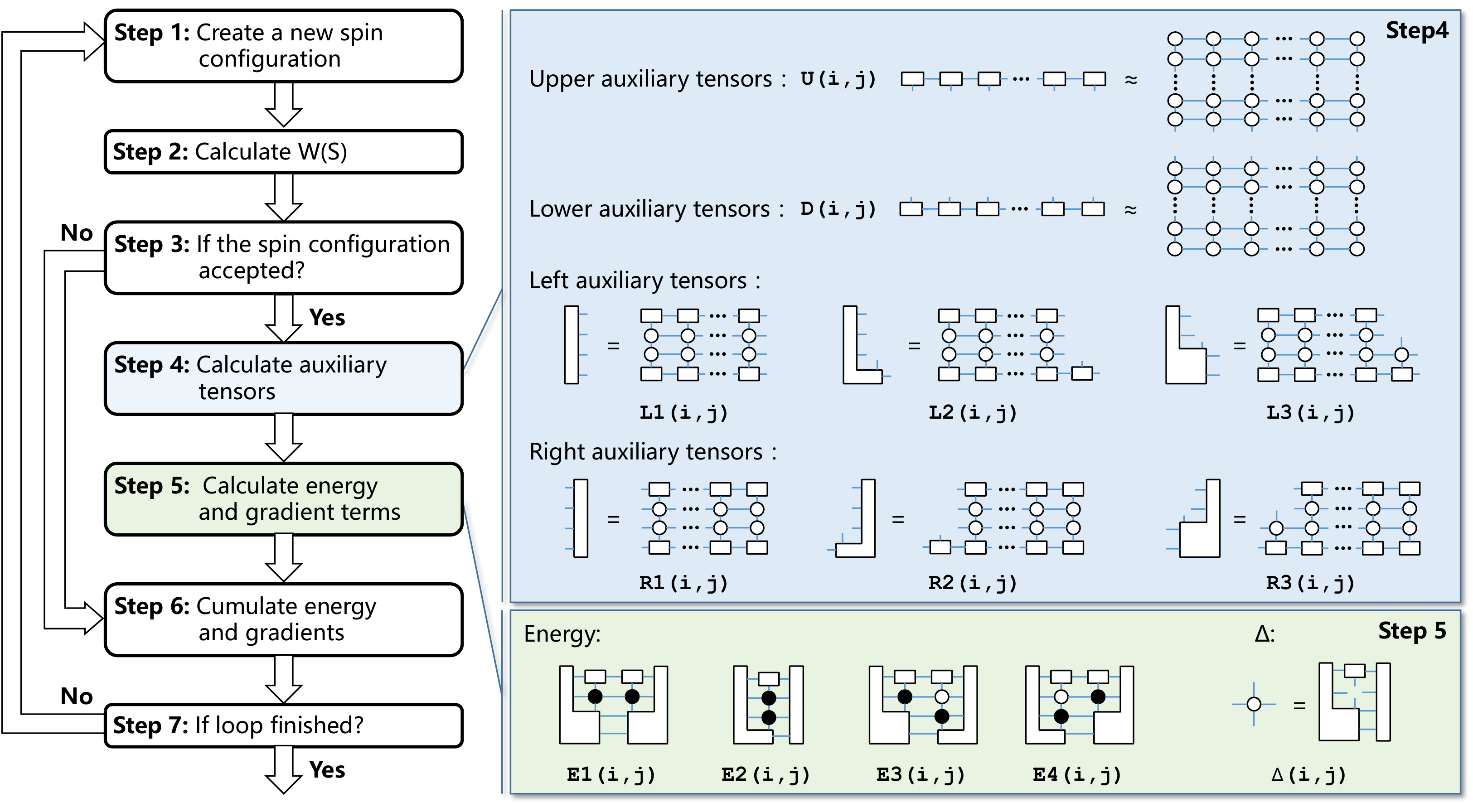}
  \caption{Thread-level parallelization of the MC sampling with tensor contractions of exotic shapes.}
  \label{fig:contraction}
\end{figure*}

In the MC sampling, the most time consuming part is to contract a single layer PEPS network for
each spin configuration $|S\rangle$, which still has rather high cost of O($D^6$).
Parallelization over this part by exploiting the thread-level concurrency
is also essential to deal with systems with large bond dimension $D$.
The parallelization of the MC sampling is shown in Figure~\ref{fig:contraction}.
We first randomly create a spin configuration (Step 1)
and then compute the coefficients $W(S)$ of the spin configuration
through single-layer tensor contraction (Step 2)
before evaluating whether it is accepted or not (Step 3).
Once it is accepted, we first calculate the exotic upper/lower/left/right auxiliary tensors (Step 4),
which will be used later to save total computing time,
and then calculate the sampling values for the energy and gradient terms (Step 5).
After that, we cumulate the energy and gradient results (Step 6).
If the spin configuration is not accepted in Step 3, we go directly
to Step 6. We then go back to the beginning the MC sampling loop Step 1.
In summary, the whole process of the MC sampling depends heavily on contracting
exotic high rank tensors, which is optimized to take advantage the on-chip parallelism;
details on how to optimize them will be shown later.

\subsection{Implementation on Sunway TaihuLight}\label{sec:optimization}

\subsubsection{Brief overview of Sunway TaihuLight}

\begin{figure}[!htb]
  \centering
  \includegraphics[width=0.48\textwidth]{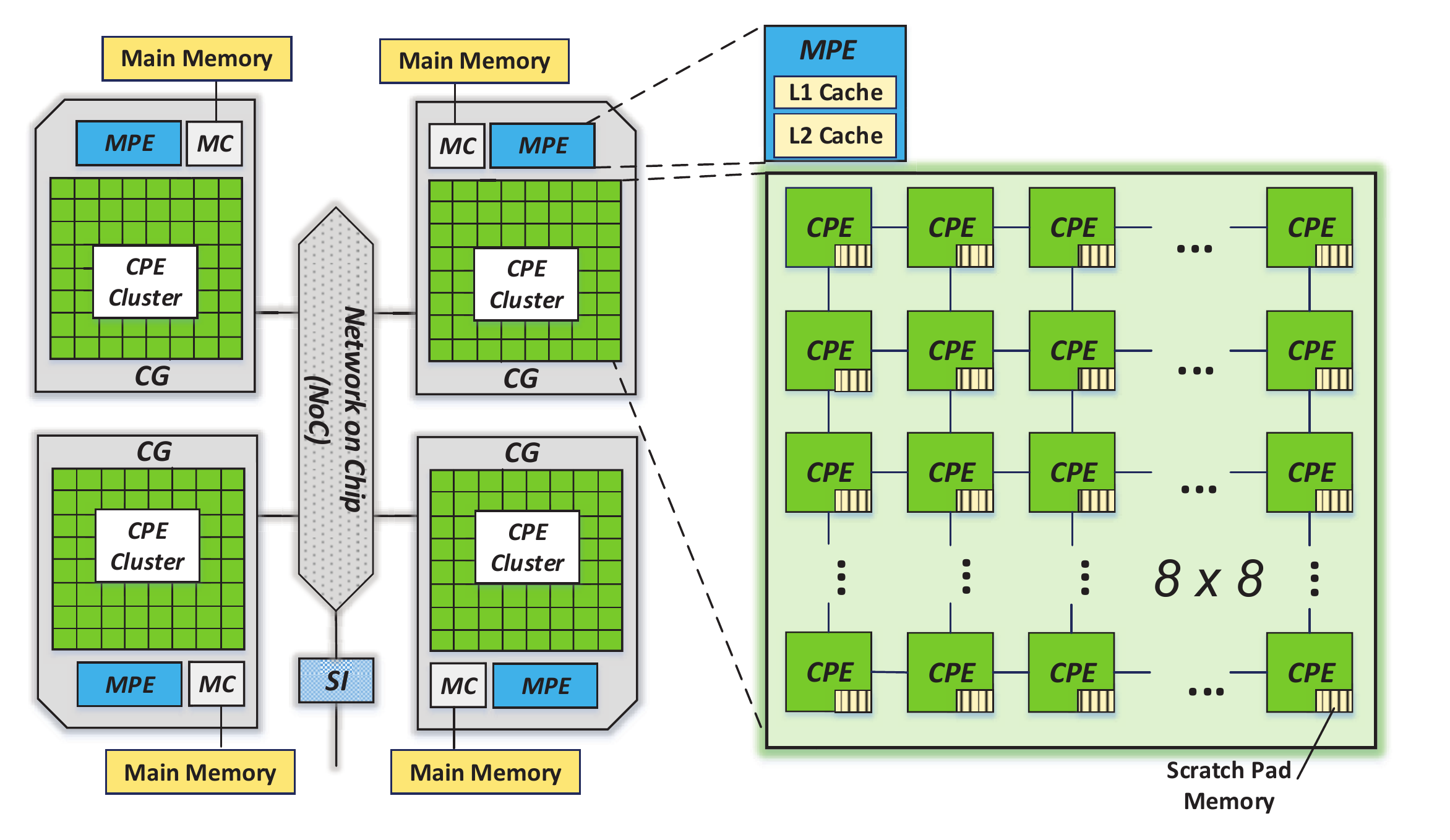}
  \caption{The SW26010 processor in Sunway TaihuLight.}
  \label{fig:sw26010}
\end{figure}

The Sunway TaihuLight supercomputer \cite{fu16_sunway,dongarra16_sunway} is built upon the 260-core SW26010 processor
that is designed as a highly integrated architecture for high performance computing \cite{zheng15_sw26010}.
Each SW26010 processor can deliver 3.06 TFLOPS performance in double precision.
The computer has 40 cabinets with each consisting of 1,024 processors,
interconnected with each other via an Infiniband network with 16 GB/s link bandwidth.
Overall, there are 40,960 such processors in the whole system, leading to an aggregate
peak performance of 125.4 PFLOPS.
%The computer has 40 cabinets, each of which hold 4 super nodes that are connected to the central network
%via 64 optical fibers, which means that the four super nodes share a fiber.
%Each super node includes 256 nodes.
%%The configuration of the Sunway TaihuLight System is listed in Table1.
%%Table 2 shows the major parameters of the TaihuLight.
%The I/O network connects 272 I/O nodes to the super node network.
%Two optical fibers dedicated to I/O are assigned to each super node.
%The link bandwidth of the network is 16 GB/s, the aggregate bandwidth is 462 TB/s,
%the bi-section bandwidth is 70 TB/s, and the I/O is 288 GB/s.
As shown in Figure \ref{fig:sw26010}, each SW26010 processor has four core groups (CGs),
connected via a low latency on-chip network. Each CG consists of one management processing element (MPE),
one computing processing element (CPE) cluster, and one memory controller (MC).
Each MPE has a 32 KB L1 data cache, a 32 KB L1 instruction cache,
and a 256 KB L2 cache. Each CPE has a 16 KB L1 instruction cache and a 64KB scratch pad memory,
which is also called local device memory (LDM).
Similar to shared memory in typical GPUs, programmers need to explicitly control the LDM to achieve high performance.
The 64 CPEs attached with a same MPE are organized as a CPE cluster.
CPEs within a cluster are connected in an 8x8 mesh topology by register buses.

\subsubsection{TNSPackage for tensor network states algorithms}

The implementation of PEPS++ method depends heavily on the operations of
high-rank tensors, including contraction, permutation, and reshape tensors, etc..
We develop a Fortran package, called TNSPackage,
which includes all kinds of basic tensor operations designed
for tensor network states (TNS) algorithms.
The PEPS++ program is implemented based on the TNSPackage, which
greatly simplifies the coding works.
An advantage of using the TNSPackage is that the thread-level
parallelism can be successfully hidden and isolated from the process-level
parallelism managed in the PEPS++ main program.
We have made great efforts to optimize and improve
the efficiency of high-rank tensor operations, by transforming the tensors to
dense matrices, so that high-performance matrix operations, such as
matrix multiplications, permutation, and QR decompositions etc. can be used
to accelerate the computation.
For example, tensor contraction is the basic and most time-consuming operation
in the PEPS++ program.
To accelerate the operation, we need to reformat the tensor contractions
into two-dimensional matrix multiplications.
To make this possible, we first rearrange the dimensions of the tensor
and then reshape the tensors to matrices.
We carefully design the optimal contraction order to reduce the numbers of permutations,
as well as the memory consumption.
In particular, on Sunway TaihuLight, the thread-level parallelism and optimization
of the time-consuming kernels mentioned above will be addressed below.

\subsubsection{Permutation optimizations}

In order to reduce the cost of high-rank tensor permutations,
we reduce them into 3D or 2D transpositions through reshaping dimensions.
They work in the out-of-place style for easy implementation and a high bandwidth utilization.
The 3D permutation kernel works in two modes {depending on whether fixing
the first or the third dimension while swapping the other two dimensions;
e.g., ($A$,$B$,$C$)$\Rightarrow$($B$,$A$,$C$), ($A$,$B$,$C$)$\Rightarrow$($B$,$A$,$C$)}.
The first mode which is preferentially used can achieve a higher memory bandwidth
than the second due to the consecutive read and write of the first dimension.
A blocked method is employed to both 3D and 2D permutations on various shape and
size of matrices. The CPEs are invoked to transpose the matrix tiles loaded
and stored with the DMA operations. For the performance difference of DMA read
and write, non-square tile shapes are chosen. The tile size is limited to half
of the LDM space so that a SIMD-based out-of-place transpose kernel can be employed.
{In addition to that, the permutation strategies are adapted to take advantage of
the matrix transposes provided in the transposed versions of DGEMM.}

\subsubsection{DGEMM optimizations}

\begin{figure}[!hbt]
  \centering
  \includegraphics[width=0.48\textwidth]{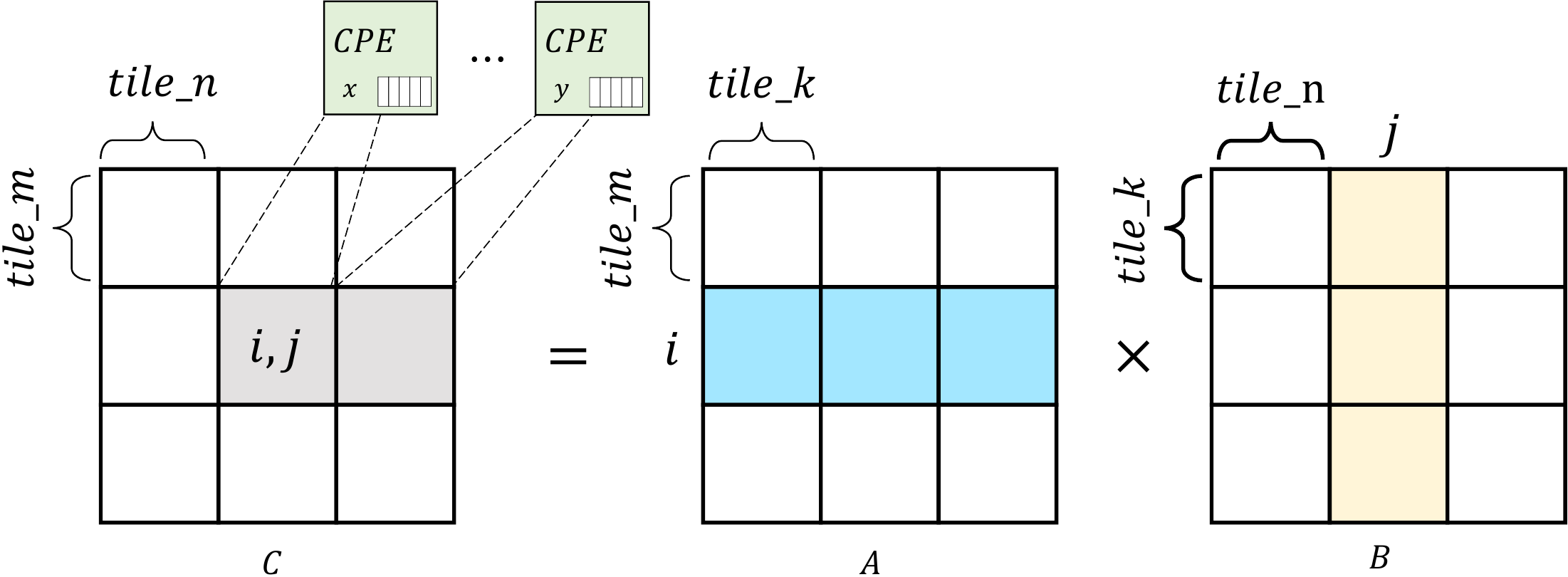}
  \caption{Performance optimizations of DGEMM.}
  \label{fig:dgemm}
\end{figure}

\begin{figure}[!hbt]
  \centering
  \includegraphics[width=0.3\textwidth]{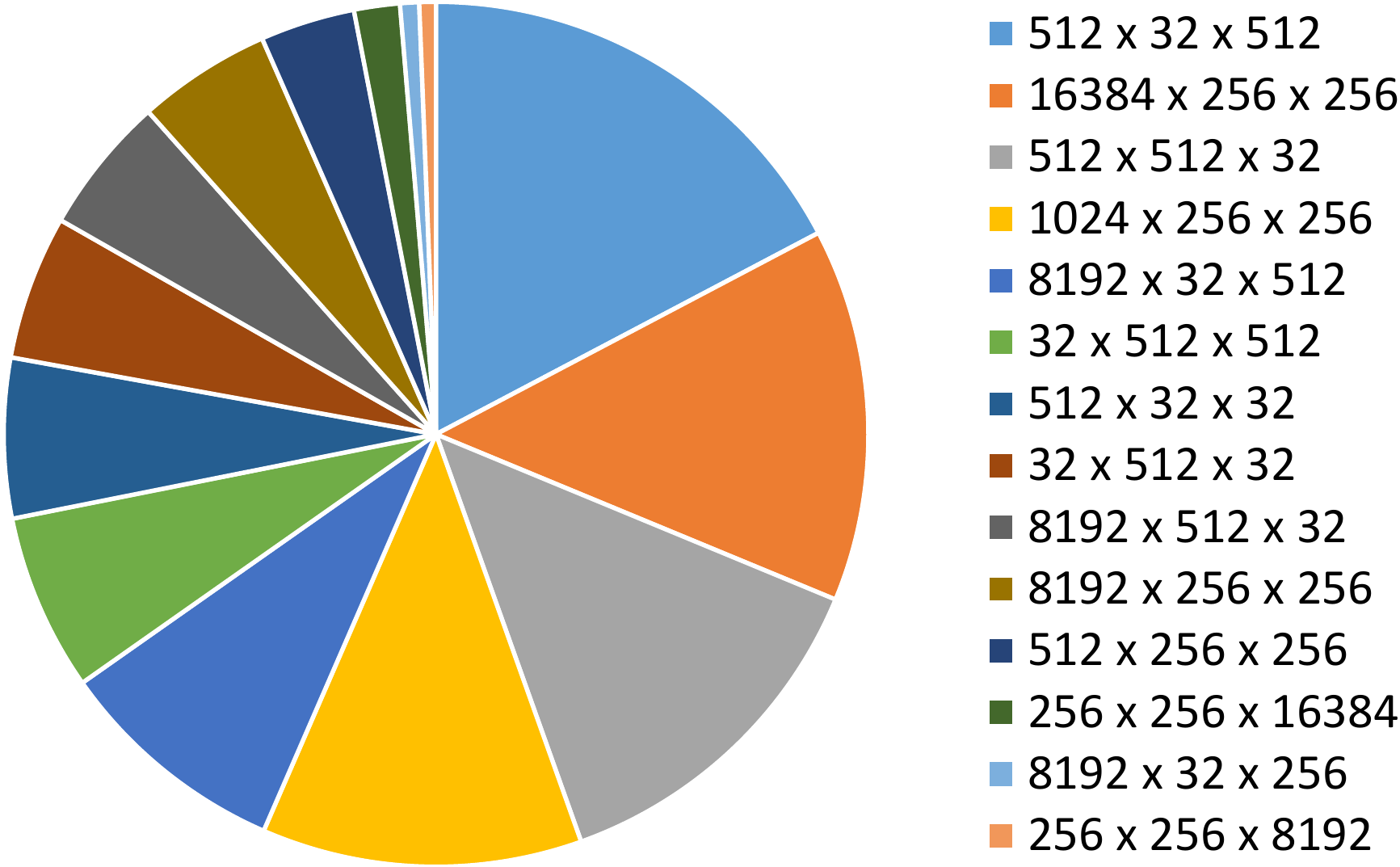}
  \caption{Various matrix sizes of DGEMM in a typical running scenario ($D=16$, $D_c=32$).}
  \label{fig:dgemm_ratio}
\end{figure}

The performance of DGEMM for matrix-matrix multiplications is critical for PEPS++.
One particular issue of DGEMM in PEPS++ is that the matrices are highly irregular
due to the intrinsic nature of the high-rank tensors.
For example, the statistics of dimension ratios of DGEMM in a typical running scenario
of PEPS++ are provided in Figure~\ref{fig:dgemm_ratio}.
To achieve high performance with various sizes of DGEMM matrices, we employ
the widely adopted parallel blocked matrix-matrix multiplication algorithm \cite{Goto2008Anatomy}
and customize it on SW26010 with further optimizations.
In the blocked DGEMM algorithm, the product of the $i$-th tile row in matrix $A$ and
the $j$-th tile column in matrix $B$ updates the resultant tile ($i$,$j$)
in matrix $C$, as shown in Figure~\ref{fig:dgemm}.
To avoid repetitive loads/stores of the tiles in Matrix $C$ and ensure parallelism,
all the CPEs are assigned to perform complete updates to these independent tiles in $C$.
On top of that, we manually arrange the SIMD assembly codes in the kernel to optimize the computing performance
and apply a dynamic work-sharing strategy reduce the imbalance caused by handling boundaries and run-time contentions.
The relevant tiles for each CPE are loaded/stored by high bandwidth asynchronized DMA mechanism
where a double-buffering technique is used to hide the non-negligible transferring time.
Performance-related parameters such as the tile size in each dimension and the number of threads
are auto-tuned to better fit with various irregular matrices.

\subsubsection{Other Optimizations}

Several global optimizations are attempted to sustain an averagely high performance in all other operations,
including matrix vector multiplications (DGEMV), QR factorizations (DGEQRF, DORGQR), vector updates (DDOT, DSCAL, DAXPY, DCOPY),
among others.
Under the circumstances of the large number of irregular data structures, it is a necessity that we implement the MPE alternatives of each primitive, so that the thread spawning overhead can be avoided by directly calling the MPE version when the input size is too small. Further, the important running parameters, such as the number of CPEs and LDM buffer sizes etc. are tuned on the basis of exhaustive empirical experiments and automatically selected during runtime. Besides, the CPEs spawned in a parallel section are  distributed to exploit all four sets of on-mesh data buses so as to maximize the bandwidth usage and avoid bus contentions. Lastly, we manually align all arrays in the code to a 256-byte which in turn can effectively enhance the DMA efficiency.

\section{How Performance Was Measured}

\begin{figure}
 \centering
 \includegraphics[width=0.45\textwidth]{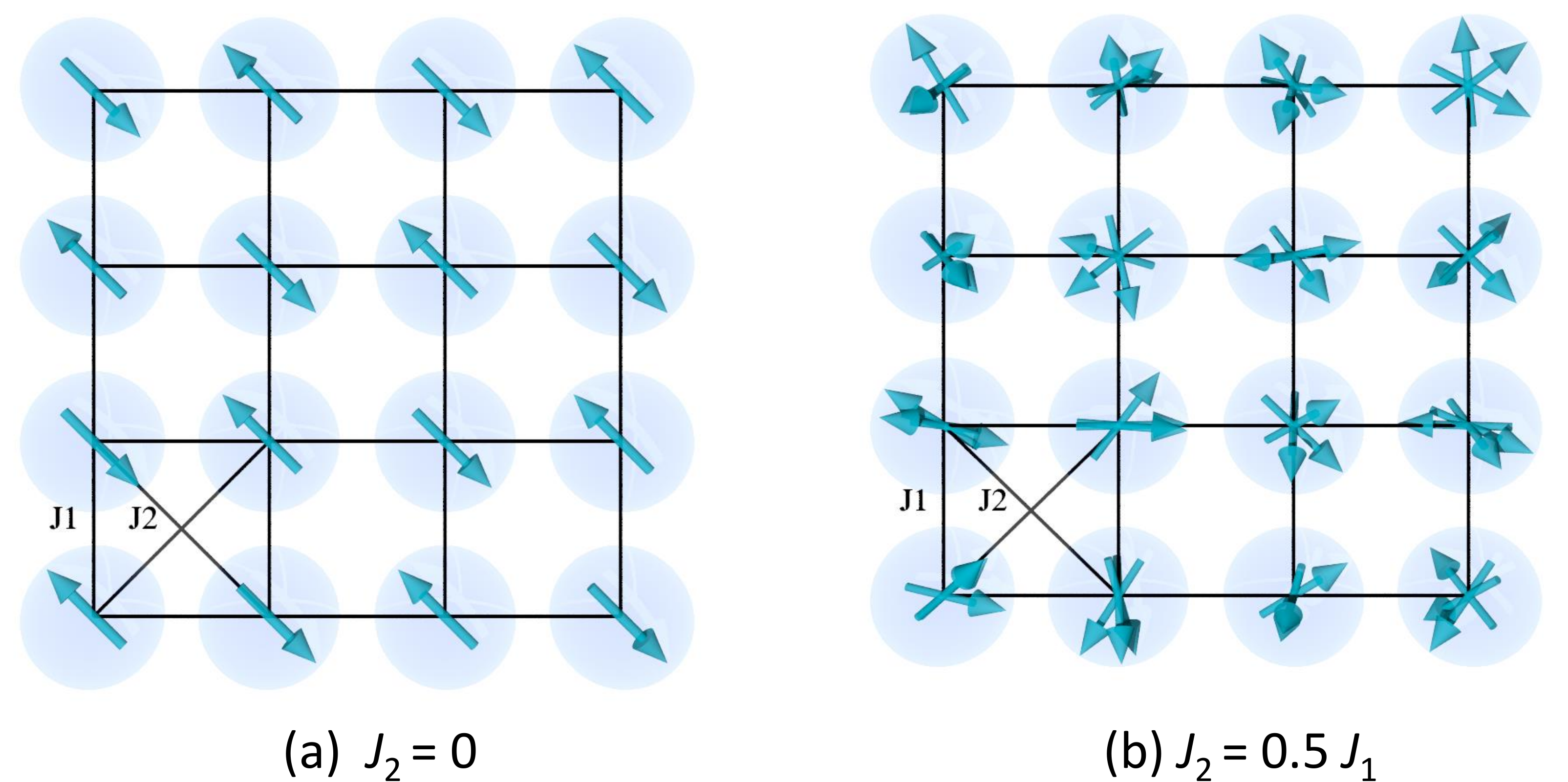}
 \caption{Schematic diagram of $J_1$-$J_2$ model and quantum spin liquid (QSL) state.
 $J_1$, $J_2$ are antiferromagnetic (AFM) exchange interactions
 between the nearest-neighbor and the next-nearest-neighbor spins, respectively.
 (a) For $J_2$=0, the ground state is dominated by the AFM spin configuration.
 (b) When $J_2 \simeq$0.5$J_1$, the ground state may be
 a superposition of massive degenerated states, known as QSL.}
 \label{fig:QSL}
\end{figure}

\subsection{Experiment Setup}

We demonstrate PEPS++ method and code by solving a strongly frustrated $J_1$-$J_2$
quantum spin model on square lattices (see Fig.~\ref{fig:QSL}).
The model is a promising candidate to search for the exotic and highly entangled
quantum spin liquid state.
The model Hamiltonian is,
\begin{equation}
H=J_1\sum_{\langle i,j \rangle}{\bf S}_i\cdot {\bf S}_j
+J_2\sum_{\langle \langle i,j \rangle\rangle} {\bf S_i} \cdot {\bf S}_j  \, ,
\end{equation}
where $\langle i,j \rangle$ and $\langle \langle i,j \rangle\rangle$ denote
the nearest neighbor (NN) and the next nearest neighbor (NNN) spin pairs respectively. We set $J_1$=1 in all experiments.
It is known that the NN interactions favorite an antiferromagnetic (AFM) order, whereas the
NNN terms favorite a stripe order. In the most frustrated region near $J_2$=0.5, because of the strong competition
between the NN and NNN interactions, a QSL may appear whereas the above two ordered states disappear.
Although the model has been studied by various methods, the results are still very controversial.

 \begin{figure}[bt]
  \centering
  \includegraphics[width=0.5\textwidth]{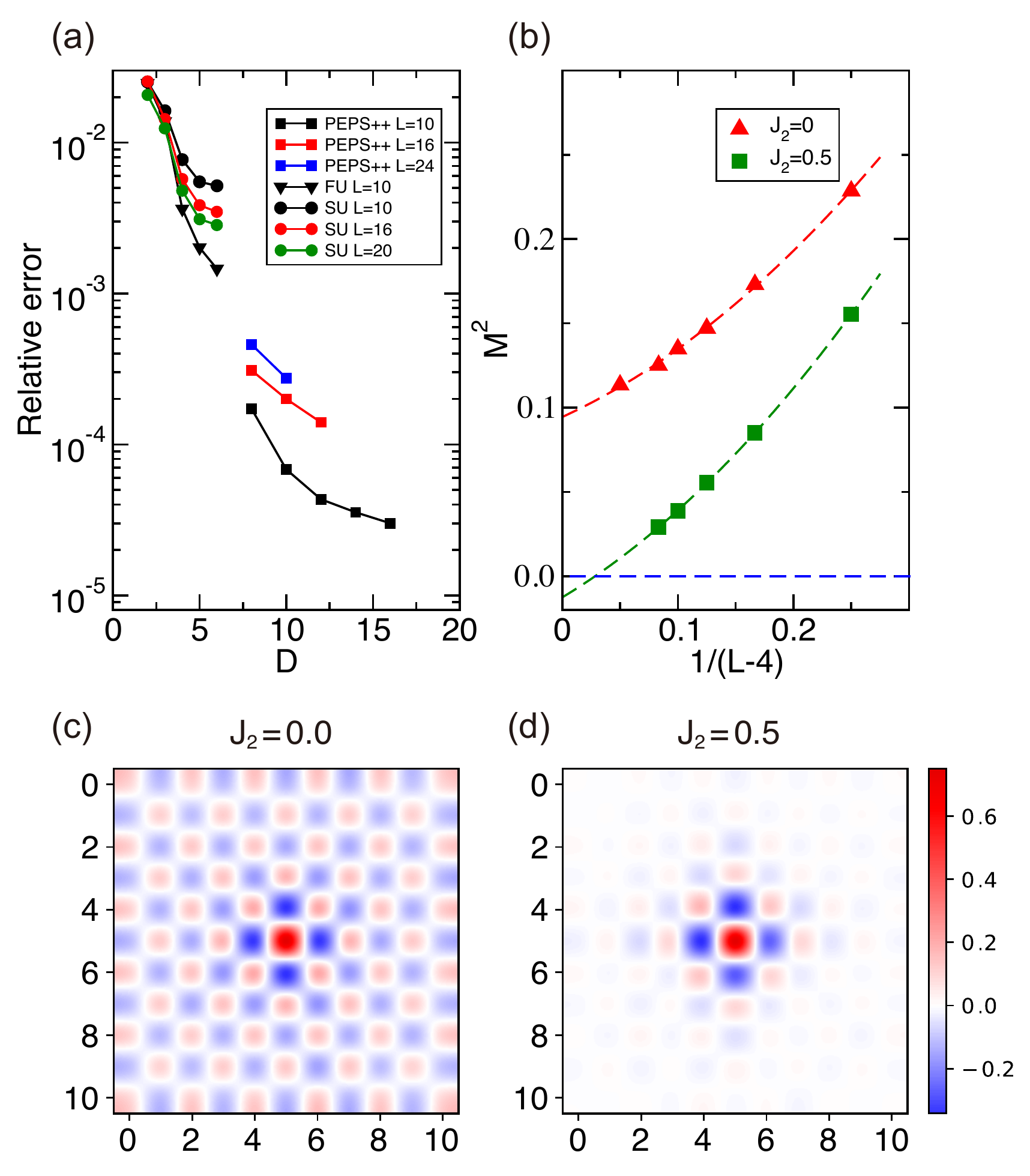}
  \caption{(a) Compare the relative errors of the SU (circle),
  FU (triangle) and PEPS++ (square) methods as functions of
    $D$ in the case of $J_2$=0 for various sizes $L$. (b) The staggered magnetization $M^2$
    for $J_2$=0 and $J_2$=0.5 as functions of 1/$(L-4)$, calculated using $D$=10.
    For $J_2$=0.5, $M^2$=0, as 1/$(L-4)$ approaching 0.
    (c) The spin-spin correlation function for $J_2$=0, showing long range
      AFM order; and (d) the spin-spin correlation function on the 2D lattice for $J_2$=0.5,
      which has only short range magnetic order, implying the ground state is a
      QSL state.}
  \label{fig:results}
\end{figure}

We first validate PEPS++ by studying the case of $J_2$=0, where
the results of QMC can be used as benchmarks.
We then study the case of $J_2$=0.5, where QMC fails due to the sign problem.
We use $D_c$=2$D$ for the $J_2$=0 case, and $D_c$=3$D$ for $J_2$=0.5, which converge
the results very well.
The maximum MC sampling number we use in the calculations is 20000 per site,
which suppresses the MC sampling error to less than 10$^{-5}$.
Figure \ref{fig:results}(a) compares the energy errors relative to the QMC results, defined as
$|E_{\rm PEPS}-E_{\rm QMC}|/E_{\rm QMC}$,
in the case of $J_2$=0, for $L_x$=$L_y$=10, 16, 24, calculated by SU, FU and PEPS++
methods. The results of SU and FU are taken from Ref.~\cite{lubasch14}.
Generally the error decreases with the increasing $D$.
The typical relative errors of SU method are approximately $\sim$10$^{-3}$ - 10$^{-2}$,
whereas the FU method gives better results. The best FU result for $L$=10 has a
relative error $\sim$10$^{-3}$, which may not be completely reliable for studying QSL.
By using PEPS++, the error is approaching 10$^{-5}$ for $L$=10
if $D$=16 is used, which is significantly
better than the best FU results.
We observe that however, the error may be lager for lager systems as shown in
Fig.~\ref{fig:results}(a). This may be because the larger systems have long correlation length,
and need significant more MC samplings, and more GO steps to converge the results. An important
work in future is to improve the MC sampling efficiency and GO convergent rate.
Since different geometries and boundary conditions are used, it is very difficult to directly
compare the energy at $J_2$=0.5 with previous state of art DMRG results~\cite{Gong14}.
Here we only list the energy on the 16$\times$16 lattice
at different virtual dimension $D$ in Table~\ref{tab:j2} as a benchmark.

\begin{table}[!ht]
\caption{Energy of $J_1$-$J_2$ model on the 16$\times$16 square lattice
at $J_2$=0.5 as a function of virtual dimension $D$ .}
\label{tab:j2}
\centering
\begin{tabular}{cccccc}
\hline\hline
 $D$ &  4 & 6 & 8 & 10 & 12  \\
\hline
Energy & -0.48885  & -0.48991  & -0.49020 & -0.49037 & -0.49051\\
\hline\hline
\end{tabular}
\end{table}

We also calculate the spin-spin correlation function
$m^2_s({\bf k})=\frac{1}{N^2}\sum_{mm'}{\langle{\bf S}_m \cdot {\bf S}_{m'}}\rangle {e}^{i {\bf k}\cdot({\bf r}_m-{\bf r}_{m'})}$,
where $N$ is the total number of spins included in the summation, using the optimized PEPS wave functions.
To reduce the boundary effects, we restrict our summation to the central lattice with
bulk size\cite{2dDMRG12} $(L-4)$$\times$$(L-4)$ to obtain $M^2(L)$ , where $N=(L-4)^2$.
We calculate $M^2$ on the lattice with different size
$L$=8 - 24. It is well known that the ground state in the case of $J_2$=0 has an AFM order,
with non-zero staggered magnetization $M^2$= $m^2_s(\pi,\pi)$.
The results are shown in Fig.~\ref{fig:results}(b).
We extrapolate $M^2(L)$  to the thermodynamic limit
$L\rightarrow \infty$ using a second-order polynomial function and
obtain $M^2(\infty)=0.094\pm0.001$, i.e., $M(\infty)$=0.307$\pm0.002$,
which is in excellent agreement with the MC result
$M_{\rm QMC}(\infty)$=0.307.
We also calculate $M^2(L)$ for $J_2$=0.5, and perform
finite size scaling. Clearly, as $L\rightarrow \infty$, we
obtain $M^2(\infty)$=0,
i.e., there is no long range magnetic order in the system,
implying the ground state of $J_2$=0 is a QSL state.

To illustrate the difference between the ground states of
$J_2$=0 and $J_2$=0.5, we plot the
spin correlations $\langle {\bf S}_{0}$$\cdot$${\bf S}_m\rangle$ on
the 2D lattice, with ${\bf S}_{0}$ placed in the center of the lattice,
in Figure~\ref{fig:results}(c),(d) respectively.
Clearly there is a long AFM order in the case of $J_2$=0, whereas for $J_2$=0.5,
there is only a short range order, but no long range order, which is consistent
with the results in Figure~\ref{fig:results}(b).
More subtle properties of the QSL states can be investigated by
various correlation functions, which can be calculated with similar complexity
as those of spin-spin correlation functions.
More detailed results and discussion on the properties of QSL in $J_1$-$J_2$ model
will be presented in a separate paper.

\subsection{Performance Measurement}
% system and environment where performance was measured (1 p max)

To perform an accurate performance measurement of our PEPS++ code on Sunway TaihuLight,
we collect the number of double-precision arithmetic operations by adding instrumentation code
in all kernel functions including, e.g., DGEMM, DGEQRF, DORGQR, DDOT, DAXPY, DSCAL.
The same code is ported to an Intel Xeon E5-2680v3 platform on which we use the PAPI performance
monitor to count total number of floating-point operations. The results obtained by the two methods
differ by less than 2\%, suggesting that our performance measurement on Sunway TaihuLight is accurate enough.

\section{Performance Results}
% include scalability (weak and strong), time to solution, efficiency (of bottleneck resources), and peak performance (2 pp max)

In our experiment, we fix the virtual bond dimension and its truncation
to $D=16$, $D_c=32$ and control two parameters including the size of the system $N=L\times L$
and the total number of samplings which is defined as the number of spin configurations accepted
during the MC sampling process.

\subsection{Many-core Acceleration on SW26010}

\begin{figure}[hbt]
  \centering
  \includegraphics[width=0.45\textwidth]{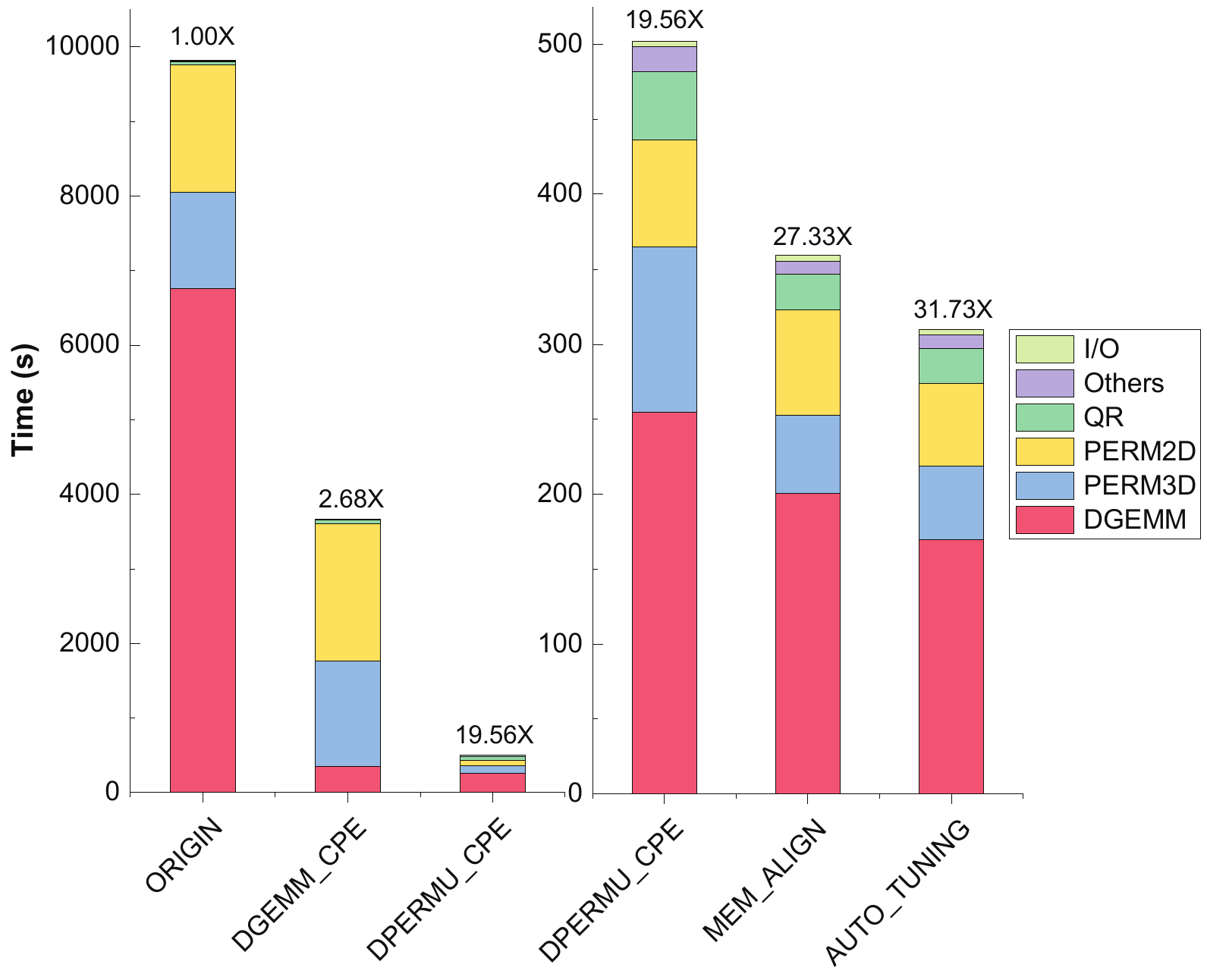}
  \caption{Performance comparison on different optimized versions of PEPS++ using a single core group. Six categories are profiled in the time breakdown, namely matrix-matrix multiplications (DGEMM), high-rank tensor permutations in 2D and 3D (PERM2D, PERM3D), QR decompositions (QR), other various operations (Others), and I/O.}
  \label{fig:Time-breakdown}
\end{figure}

First we evaluate the performance of PEPS++ on a single core group to for a problem of $N=16\times16$
to examine the effects of different optimization techniques; the results are shown in Figure~\ref{fig:Time-breakdown}.
We run the test and record the total execution time of ten gradient optimization steps.
The baseline we choose is the a highly optimized MPE-only version of PEPS++
linked with the MPE version of the xMath extended math library.
From the figure we can see that the performance improvement is remarkable when
applying different optimization techniques on the CPEs as described in Section~\ref{sec:optimization},
including the DGEMM optimization, the DGEMM-aware permutation strategy, the tensor data alignment,
and the auto-turning method. A breakdown of the runtime is also provided in the figure for further analysis.
The DGEMM operations contribute around 50\% of the execution time and the permutations cost another 30\%.
In particular, the average performance of DGEMM on the various irregular shaped matrices
can deliver over 300 GFLOPS performance and the memory bandwidth usage of the 2D and 3D permutations
can reach 17.6 GB/s and 19.5 GB/s, respectively.
With all optimizations utilized, PEPS++ can sustain a final speedup of 31.73X as compared
to the optimized MPE-only version.

\subsection{Strong Scaling Results}

\begin{table}[!ht]
\caption{Configuration of strong scaling tests.}
\label{tab:strong}
\setlength\tabcolsep{5.5pt}
\centering
\begin{tabular}{cccc}
\hline
\# Processes& \# Sites& \# Samplings/Process & \# Samplings \\
\hline
%168~$\times$~~38~$\times$~1&16128~$\times$~~2432~$\times$~128&~32256~$\times$~~4864~$\times$~128 \\
   {\,\,10,000} & 16 $\times$ 16 & {256} & {2,560,000} \\
   {\,\,18,000} & 16 $\times$ 16 & {142} & {2,556,000} \\
   {\,\,31,000} & 16 $\times$ 16 & {\,\,83} & {2,573,000} \\
   {\,\,57,000} & 16 $\times$ 16 & {\,\,45} & {2,565,000} \\
   {100,000} & 16 $\times$ 16 & {\,\,26} & {2,600,000} \\
   {160,000} & 16 $\times$ 16 & {\,\,16} & {2,560,000} \\
\hline
\end{tabular}
\end{table}

In the parallel scalability tests, we record the total execution time of a single gradient optimization step.
In the strong scaling tests, we increase the total number of MPI processes,
with each corresponding to one core group of 65 cores,
to investigate the overall speedup of solving a large problem with fixed problem size.
The detailed configuration of strong scalability tests is shown in Table~\ref{tab:strong}.
In the tests, the number of sites is fixed to be $N=16\times16$
and the number of samplings per process is decreased in proportion to
the number of processes so that the total number of samplings is kept around
$2,560,000$.

\begin{figure}[hbt]
  \centering
  \includegraphics[width=.86\columnwidth]{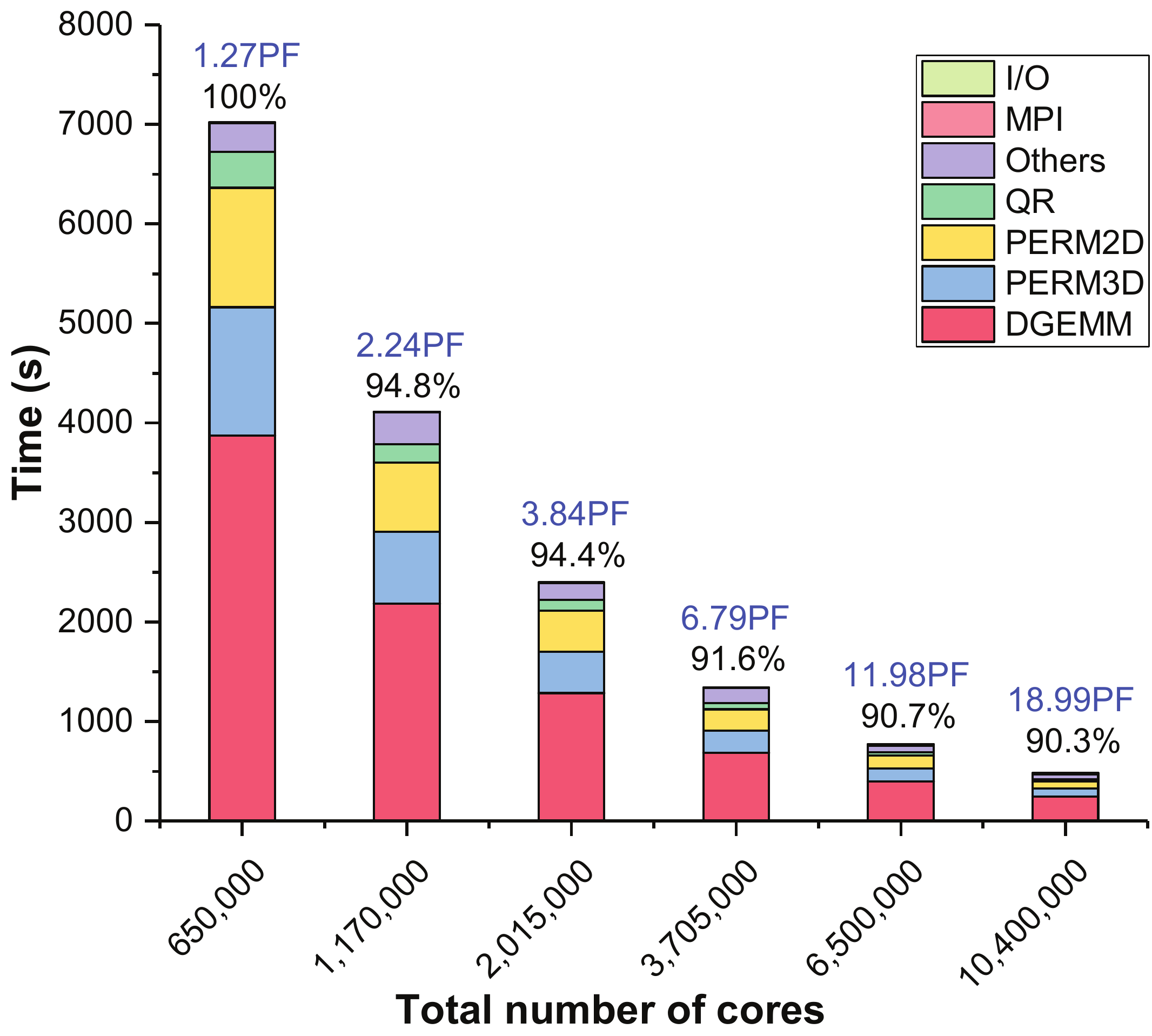}
  \caption{Strong scaling results from 650,000 cores (10,000 core groups) to 10,400,000 cores (160,000 core groups). }
  \label{fig:strong}
\end{figure}

We run the test and record the total execution time of a single gradient optimization step.
The test results are presented in Figure~\ref{fig:strong}, in which
the parallel efficiency, the sustained PFLOPS, and the total runtime are also summarized.
From the figure, it is observed that the major cost of PEPS++ is
the DGEMM on various irregular matrices and the 2D and 3D permutation of high-rank tensors,
contributing together around 82\% to 91\% of the total runtime.
Besides that, the I/O and MPI communication takes only a negligible portion
of the total runtime. The other operations including
QR factorizations, vector BLAS-1 updates, etc. contribute around 7\%.
Overall it shows that our implementation and optimization of PEPS++
on Sunway TaihuLight can dramatically increase the performance
of all major operations with more processors in use.
In terms of the strong scalability, PEPS++ can successfully reduce the
time-to-solution by 14.88 fold when the total number of processor cores
is increased from 65,000 to over 10 million with a parallel efficiency
of over 90\% and an aggregate performance of 18.99 PFLOPS
in double precision.

\subsection{Weak Scaling Results}

\begin{table}[!ht]
\caption{Configuration of weak scaling tests.}
\label{tab:weak}
\setlength\tabcolsep{5.5pt}
\centering
\begin{tabular}{cccc}
\hline
\# Processes& \# Sites& \# Samplings/Process & \# Samplings \\
\hline
   {\,\,\,\,8,000} & 10 $\times$ 10 & {80} & {\,\,\,640,000} \\
   {\,\,18,000} & 12 $\times$ 12 & {51} & {\,\,\,921,600} \\
   {\,\,34,000} & 14 $\times$ 14 & {36} & {\,\,1,254,400} \\
   {\,\,60,000} & 16 $\times$ 16 & {26} & {\,\,1,638,400} \\
   {100,000} & 18 $\times$ 18 & {20} & {\,\,2,073,600} \\
   {160,000} & 20 $\times$ 20 & {16} & {\,\,2,560,000} \\
\hline
\end{tabular}
\end{table}

Different from the strong scaling run, the purpose of the weak scaling tests is
to examine the performance of solving large problems when more processor cores are used.
Considering the boundary effects, the contributed degree
of freedom of the lattice is in the order of $(L-2)^2$ when the number
of sites is $N=L\times L$. While the number of sites  increased, the
number of sampling should be increased accordingly. Based on the above
principles, we design the weak scaling tests with the configuration
listed in Table~\ref{tab:weak}. In the tests both the number of sites
and the number of samplings are adjusted as the number of MPI processes
changes so that the averaged degree of freedom per process is maintained
around 5,120.

\begin{figure}[hbt]
  \centering
  \includegraphics[width=.98\columnwidth]{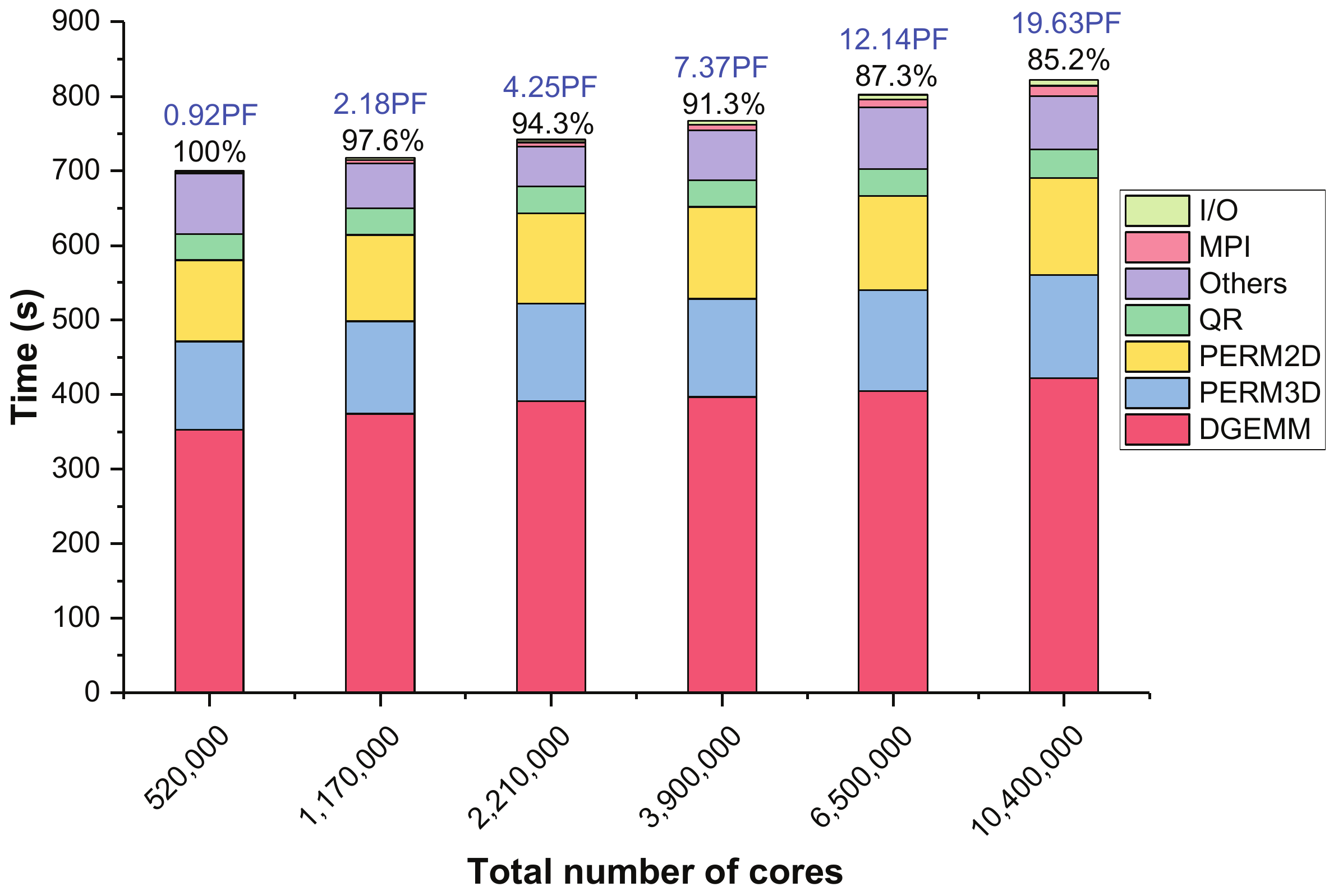}
  \caption{Weak scaling results from 520,000 cores (8,000 core groups) to 10,400,000 cores (160,000 core groups).}
  \label{fig:weak}
\end{figure}

The weak scaling results are drawn in Figure~\ref{fig:weak}. Some additional information
on the parallel efficiency, the sustained PFLOPS, and the total runtime are also given.
Again, we can observe that DGEMM and tensor permutations takes a major part of the total
runtime, while I/O cost is negligible. Unlike the strong scaling results, the
MPI communication in the weak scaling tests takes larger time, but is still below
2\% of the total. In terms of weak scalability, the PEPS++ scales from 520,000 cores
to over 10 millions cores
with a parallel efficiency of 85.2\% for a 20-fold increase of problem size.
At the full system scale, the sustained performance is 19.63 PFLOPS in
double precision, which is above 16\% of the peak performance.

\section{Concluding Remarks}

The study of strongly correlated quantum systems
poses great challenges in condensed matter physics \cite{Anderson1972}.
Advanced numerical methods that can fit well on modern supercomputers have
become a key to unveiling the fascinating phenomena of strongly correlated quantum systems.
In this work, we developed PEPS++, a low scaling and highly scalable method that hybridizes
PEPS with MC sampling techniques, so as to efficiently simulate challenging problems
arising from strongly correlated quantum systems.
We demonstrated our method on a highly frustrated 2D $J_1$-$J_2$ model
and showed that the code can scale to the full system of over ten million cores on
the Sunway TaihuLight supercomputer, sustaining $\sim$20 PFLOPS performance in double precision.
With this ability, we are able to investigate systems with size up to 24$\times$24,
and bond dimension $D$ up to 16, with extremely high accuracy approaching 10$^{-5}$
and dramatically reduced time-to-solution.

In order to achieve high performance on the Sunway TaihuLight supercomputer,
we have exploited the intrinsic process-level parallelism by distributing the MC sampling
tasks across different MPI processes, and made best utilization of the on-chip
thread level parallelism by breaking down each MC sampling task into
high-rank tensor computations. Highly optimized tensor computation
kernels for the exotic tensor contraction, permutation, and reshape etc. have been
encapsulated in a standalone software package TNSPackage with both general considerations
for invoking fast BLAS and LAPACK libraries and specific treatments
for enhanced data layout and reduced memory footprint.
The implementation and optimization technique we applied in this study
may serve as a demonstration on how to deploy a tensor-based
application on Sunway TaihuLight with extreme-scale parallelisms.
Moreover, the major ideas of PEPS++ have a broader range of applications than already demonstrated in this paper, and can be generalized to a larger variety of many-core based supercomputers, including systems equipped with GPU accelerators or Intel Xeon Phi coprocessors.

With the faithful and scalable simulation results achieved in this study,
we can make a confident conclusion about the physics, especially, whether there is a spin liquid
in the strong frustrated region and
reveal some major characteristics of the highly entangled quantum system.
This research may have successfully opened up a promising route to solving various
important strongly correlated quantum many-body problems.
The PEPS++ method can be directly applied to investigate other strongly frustrated
boson and quantum spin models,
including the Kitaev's models and the Kagome model, etc.,
which are very difficult to solve by previous methods.
In addition, it can be further applied to other challenging problems,
such as the strongly correlated electronic systems for understanding the high Tc superconductivity
and the highly entangled fermion systems that is beyond the entanglement area law.


\ifCLASSOPTIONcompsoc
  \section*{Acknowledgments}
\else
  \section*{Acknowledgment}
\fi

The authors would like to acknowledge the contributions of Xin Liu for suggestions on performance optimizations.
Special thanks go to NSCC-Wuxi for providing free access to Sunway TaihuLight.
We also thank Ziyang Meng, Guangyu Sun, Binghui Yan, Chenzhi Liao, Hao Zhang, Yang Yu, Zhichao Su, and Lijuan Jiang for
valuable technical discussions. The work is supported by the National Key Research and Development Program of China
(Grants No. 2016YFB0201202 and 2016YFB0201902) and National Natural Science Foundation of China (Grant No. 91530323).
The corresponding author is Chao Yang (chao\_yang@pku.edu.cn).

\IEEEtriggeratref{30}


%\bibliographystyle{IEEEtran}
%\bibliography{gbprize}


\end{document}